\documentclass[aps,floats,twocolumn,epsf,prb,showpacs]{revtex4-1}
\usepackage{graphicx}
\usepackage{epstopdf}
\usepackage{amsmath}
\usepackage{amssymb}
\usepackage[colorlinks=true,urlcolor=blue,citecolor=blue,linkcolor=blue,breaklinks=true]{hyperref}
\usepackage{color}
\usepackage[dvipsnames]{xcolor}
\usepackage{ulem}   

\newcommand{\mk}[1]{{\textcolor{black}{#1}}}

\begin{document}

\title{Natural orbital impurity solver for real-frequency properties at finite temperature}

\author{Motoharu Kitatani$^{a,b}$, Shiro Sakai$^a$, Ryotaro Arita$^{a,c}$}

\affiliation{$^a$RIKEN Center for Emergent Matter Science (CEMS), Wako, Saitama, 351-0198, Japan}
\affiliation{$^b$Department of Material Science, University of Hyogo, Ako, Hyogo 678-1297, Japan}
\affiliation{$^c$Research Center for Advanced Science and Technology, University of Tokyo 4-6-1, Komaba, Meguro-ku, Tokyo 153-8904, Japan}

\date{\today}

\begin{abstract}
We extend the natural orbital impurity solver [Y. Lu, M. H\"oppner, O. Gunnarsson, and M. W. Haverkort, Phys. Rev. B \textbf{90}, 085102 (2014)] to finite temperatures and apply it to calculate \mk{spectral and} transport properties of correlated electrons within the dynamical mean-field theory. First, we benchmark our method against the exact diagonalization result for small clusters, finding that the natural orbital scheme works well not only for zero temperature but for low finite temperatures. The method yields smooth and sufficiently accurate spectra, which agree well with the results of the numerical renormalization group.
Using the smooth spectra, we calculate the electric conductivity and Seebeck coefficient for the two-dimensional Hubbard model at low temperatures which are in the scope of many experiments and practical applications. 
These results demonstrate the usefulness of the natural orbital framework for obtaining the real frequency information of correlated electron systems.
\end{abstract}

\maketitle

\section{Introduction}
Understanding correlated electron systems is one of the central issues in condensed matter physics. To address the issue, the dynamical mean-field theory (DMFT)\cite{Metzner1989,Georges1992,Georges1996} has been developed and extended in its scope in the past decades.
The DMFT properly takes into account the temporal fluctuations which are essential to describe the Mott metal-insulator transition while it ignores the spatial fluctuations which could be important for low-dimensional fluctuating systems such as cuprates. Combined with {\it ab initio} calculations, the DMFT has also been used to calculate the electronic structure of real materials, where several orbital degrees of freedom are usually involved\cite{Kotliar2006,Held2007}. 

In the DMFT, we need to solve a quantum impurity model which is usually simpler than the original lattice model, but still difficult to solve analytically. Many numerical methods have been proposed for solving that model.
The continuous-time quantum Monte Carlo (CT-QMC) method \cite{Gull-RevModPhys.83.349} is one of the most commonly used methods since it gives a numerically exact result within a statistical error and is applicable to relatively large degrees of freedom. However, because it is formulated on the imaginary (Matsubara) time axis, an analytic continuation of the numerical data with statistical errors, which is an ill-posed inverse problem, is required to obtain real-frequency properties. While the numerical analytic continuation method has been improved \cite{Gubernatis91,Otsuki17,Fei21}, the dependence on the data precision is unavoidable and it is difficult to correctly describe a high-frequency structure or sharp structure like the Drude peak at low temperatures.

The real-frequency data is, however, indispensable to compare with spectroscopic experiments like optics, photoemission and scanning tunneling spectroscopies, as well as electronic Raman and resonant inelastic X-ray scatterings. Besides these, it is also necessary for calculating transport properties like electric and thermal conductivities, as well as a thermoelectric effect measured by the Seebeck or Nernst coefficient. Furthermore, the real-frequency structure of the Green function and \mk{the} self-energy can give crucial information on \mk{the} underlying mechanism of physical phenomena, as exemplified by the electron-phonon mechanism of conventional superconductivity established through the analysis of the frequency-dependent superconducting gap function \cite{MCMillan-Rowell-PhysRevLett.14.108}.

Thus, to apply the DMFT to these issues of real materials, we need an impurity solver which allows a direct access to real-frequency properties and is potentially applicable to multiorbital systems.  One way in this direction is to improve the numerical renormalization group (NRG) \cite{Bulla99,Bulla-RevModPhys.80.395} or density-matrix renormalization group (DMRG) method \cite{White92,Hallberg06,Schollwock11} for multiorbital systems, as indeed has been done in Refs.~\onlinecite{PRuschke-Bulla-EPJ.44.217, Peters-Pruschke-PhysRevB.81.035112, Stadler-PhysRevLett.115.136401,Kugler2020} up to three orbitals. Recently, the fork tensor-product state was also proposed and applied to a three orbital system\cite{PhysRevX.7.031013}.

Another way is to improve the exact diagonalization (ED) \cite{Caffarel-Krauth-PhysRevLett.72.1545} method. The problem in the ED \mk{solver} is a \mk{severe limitation in the}
number of the \mk{tractable} bath sites, which results in a discrete spectrum of the impurity Green function. Even with modern computers, it is difficult to deal with more than 20 degrees of freedom (except for spin) including both impurity and bath sites. Because of this limitation, it has been difficult to apply the ED solver to more than four site or orbital degrees of freedom \mk{at the impurity}. Also, to mitigate the finite-size effect, the spectrum is usually calculated with an energy-smearing factor (i.e., the imaginary part of the energy), which defines the width of each discrete \mk{spectral} peak. While this does not alter the position \mk{and weight} of the spectral peaks, it changes the broadness of the spectra and can considerably affect the values of transport coefficients, making it difficult to determine them unambiguously. 

Recently, an interesting idea was proposed in Refs.\onlinecite{Lu2014,Lu2019} to remedy this problem of the finite bath. Representing the bath sites in the natural orbital basis, which is defined as a basis diagonalizing the bath density matrix, Lu {\it et al.} constructed efficient algorithms to take account of low-energy excitations beyond the size tractable with the conventional ED. They calculated  a real-frequency Green function in a restricted Hilbert space, where the significant reduction of its dimension allowed them to consider $\sim 1000$ bath sites, to yield a sufficiently smooth spectrum. Considering the tractable size of the Hilbert space in the ED, the improvement by the natural orbitals may enable us to deal with up to five orbitals in near future, where many correlated materials -- in particular transition-metal compounds with five $d$ orbitals -- await the clarification by theory.

The idea of the natural-orbital solver \mk{[c.f., Fig.~1(a)]} is in line with the configuration interaction (CI) method which has recently been applied to the DMFT by several groups \cite{Zgid12,Lin13,Go15,Go17}.
The CI method is widely used for molecules in quantum chemistry where it is \mk{justifiable to use}
fully filled localized molecular orbitals and the mixture of Slater determinants \mk{only} for the rest degrees of freedom. 
On the other hand, in metallic states in solids, orbitals are more hybridized with each other and it is
\mk{intractable with}
the CI method. Actually, the ground state of the Hubbard cluster is quite entangled and difficult to describe \mk{solely} by a few Slater determinants. Nevertheless, for the quantum impurity model, it turned out that the ground state is well described by a small number of Slater determinants and low-energy particle-hole excitations from it, \mk{so that the CI method works well\cite{Lin13}}. The natural orbital framework gives us a general way to construct the new basis set where the CI scheme works efficiently. Recently, it is suggested that the natural orbital configuration is also useful for tensor network calculations\cite{cao2021tree}.

The application of the natural orbital method has, however, been limited to zero temperature. In view of the comparison with experiments, its extension to finite temperatures is indispensable.
In this paper, we attempt such an extension, by constructing different natural orbitals for each electron-spin number sector. We find that the minimum energy of each sector and the Green function are well approximated by the scheme at low but finite temperatures. 
We benchmark the accuracy of our result for a
small cluster against a direct ED calculation and show that a smooth spectrum consistent with the NRG result in the literature is obtained with a large number of bath sites.
Then, we calculate transport properties for the two-dimensional Hubbard model. 
We obtain consistent results with the previous CT-QMC+Pad\'{e} studies while the current scheme can potentially reach much lower temperatures for general models, which are relevant to many experiments and practical applications.

The paper is organized as follows: In Sec.~II, we give an overview of the natural orbital framework and introduce our extension for different sector calculations toward finite temperatures. In Sec.~III, we present some benchmark results for small clusters and large systems at zero temperature. Then, we show our results for conductivity and Seebeck coefficient for the two-dimensional Hubbard model on a square lattice. 

\section{Formalism}
We apply the idea of natural orbitals \cite{Lu2014,Lu2019} to the finite-temperature ED solver \cite{Caffarel-Krauth-PhysRevLett.72.1545,Liebsch12,Capone07} of the DMFT. In the ED, we first divide the Hilbert space into independent sectors specified by quantum numbers, instead of dealing with the whole Hilbert space at once. In this paper, we assume a spin-conserved system, so that $(n_{\uparrow},n_{\downarrow})$ specifies each sector, where $n_\sigma$ is the number of spin-$\sigma$ electrons in the whole system. We then use the Lanczos algorithm for each sector. \mk{Here, we employed ED since it is the most straightforward and controllable, while there are other numerical methods to analyze the simplified natural orbital models \cite{Kohn2021}.}

We basically follow the previously proposed algorithm of the natural orbital solver with the projection approach for calculating the Green function \cite{Lu2019}. The new point here is that not only the ground state but also all sectors contribute to the Green function at finite temperatures. For this \mk{reason},
we construct different natural orbitals for each sector to evaluate the energy difference between each sector accurately. 


\mk{In this study, we divided the entire Fock space into several sectors based on $N$ (total number of electrons) and $S_z$ ($S$: total spin), which are conserved. In the presence of the spin-orbit coupling, we can divide the space based on $N$ and $J_z$ ($J$: total angular momentum) instead. This separation is still valid in the presence of the isotropic Hund's coupling. For more general interactions, we may be able to separate the space based only on $N$, so that the dimension of each sector becomes large. This may limit the tractable size of the ${\cal H}_{\rm I}$ (i.e., the core part of the impurity Hamiltonian) though there can be other conserving quantities in some cases \cite{Parragh2012}.} In this paper, we focus on the single band model. The extension to the multi-orbital systems would be straightforward, see, e.g., Ref.~\onlinecite{cao2021tree}.

\begin{figure}[htbp]
        \centering
                \includegraphics[width=\linewidth,angle=0]{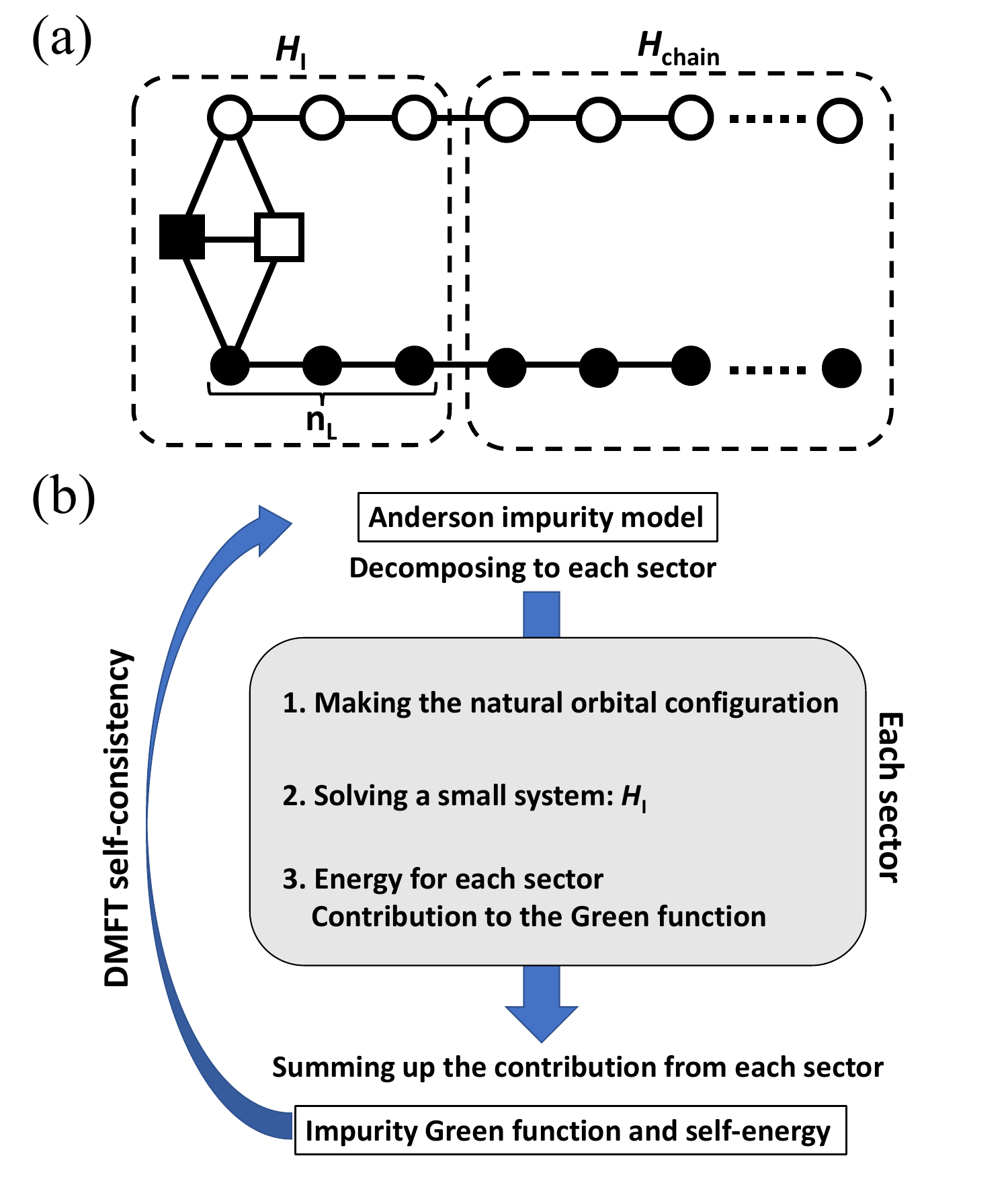}
        \caption{\mk{(a) Schematic figure of the natural orbital Hamiltonian. The dark square is the interacting impurity site and the open square is an active bath site. Filled (open) circles are valence (conduction) bath sites that are almost occupied (empty). We perform exact diagonalization for the ${\cal H}_{\rm I}$ part, and consider only a few particle excitations for ${\cal H}_{\rm chain}$. (b)} General flow of the finite temperature natural orbital impurity solver.
        \label{Fig:schematic}}
\end{figure}

The formalism is summarized in Fig~\ref{Fig:schematic}\mk{(b)}. 
We start with the standard Anderson impurity model
\begin{align}
    {\cal H}_{\rm AIM} =& {\cal H}_{\rm loc}+{\cal H}_{\rm bath}, \\
    {\cal H_{\rm loc}} =& \sum_{\sigma} \epsilon_{\rm imp} a^{\dag}_{{\rm imp},\sigma}a_{{\rm imp},\sigma}
                +U \mk{a^{\dag}_{{\rm imp},\uparrow}a_{{\rm imp},\uparrow}a^{\dag}_{{\rm imp},\downarrow}a_{{\rm imp},\downarrow}}, \\
    {\cal H}_{\rm bath}=&\sum_{i=1,\sigma}^{n_{\rm bath}} \epsilon_{i,\sigma} a^{\dag}_{i,\sigma} a_{i,\sigma}
               +\sum_{i=1,\sigma}^{n_{\rm bath}} V_i (a^{\dag}_{{\rm imp},\sigma}a_{i,\sigma} + h.c.)
               \label{Eq:bath-ham},
\end{align}
where $\epsilon_{\rm imp}$ is the energy level of the impurity site, and $a^{\dag}_{{\rm imp},\sigma}(a_{{\rm imp},\sigma})$ and $a^{\dag}_{i,\sigma}(a_{i,\sigma})$, respectively, are the electron creation (annihilation) operators at the impurity site and the $i$-th bath site with spin $\sigma \in \{\uparrow,\downarrow\}$. $U$ is the onsite Coulomb repulsion, 
and $V_i$ is the hybridization to the $i$-th bath site.

First, we choose a sector ($n_{\uparrow},n_{\downarrow})$. Then, following Ref.~\onlinecite{Lu2019}, we make the mean-field Hamiltonian $({\cal H}_{\rm MF})$ whose energy level is adjusted to keep the density at the impurity site. We calculate the density matrix with the lowest-energy state of ${\cal H}_{\rm MF}$ within the sector ($n_{\uparrow},n_{\downarrow})$.
Natural orbitals are obtained as the basis set diagonalizing the bath density matrix. Notice that even if we assume the paramagnetic system, we need to construct spin dependent natural orbitals to calculate the contribution from the spin imbalanced sector\mk{s}. After that, we construct the natural orbital representation of the quantum impurity Hamiltonian $({\cal H}^{\rm NO}_{\rm AIM})$ for each sector in the same way as the previous study in Refs.~\onlinecite{Lu2014,Lu2019}.

The natural orbital Hamiltonian contains the impurity site (imp), active bath site (b), and valence (v)/ conduction (c) chains \mk{[Fig.~1(a)]}. We then determine the cutoff length $n_{\rm L}$ for the ED calculation. Outside this cutoff, we assume the fully occupied (empty) configuration for the valence (conduction) chain. 
Thus, the whole Hamiltonian is decoupled into two terms as
\begin{align}
    {\cal H}_{\rm AIM}^{\rm NO} 
    =& {\cal H}_{\rm I} + {\cal H}_{\rm chain}, \\
    {\cal H}_{\rm I} 
    =&
    {\cal H}_{\rm loc}+\sum_{\sigma} \left[
    \epsilon_b b^{\dag}_{\sigma}b_{\sigma} 
    +V^{\mk{\rm imp,}b}_{\sigma}(b^{\dag}_{\sigma} a_{{\rm imp},\sigma} +h.c.) \right. \nonumber\\
    &\mk{+V^{{\rm imp},c}_{\sigma}(c^{\dag}_{1,\sigma} a_{{\rm imp},\sigma} +h.c.)
    +V^{{\rm imp},v}_{\sigma}(v^{\dag}_{1,\sigma} a_{{\rm imp},\sigma} +h.c.)} \nonumber \\
    &\mk{+\epsilon^{c}_{1,\sigma}
    c^{\dag}_{1,\sigma}c_{1,\sigma}
    +V^{c}_{1,\sigma}
    (c^{\dag}_{1,\sigma}b_{\sigma} +h.c.)} \nonumber \\
    &\mk{\left. +\epsilon^{v}_{1,\sigma}
    v^{\dag}_{1,\sigma}v_{1,\sigma}
    +V^{v}_{1,\sigma}
    (v^{\dag}_{1,\sigma}b_{\sigma} +h.c.)\right]} \nonumber \\
    &+
    \sum_{\sigma,i=\mk{2}}^{n_{\rm L}} 
    \left[\epsilon^{c}_{i,\sigma} c^{\dag}_{i,\sigma} c_{i,\sigma}
    +V^{c}_{i,\sigma}
    (c^{\dag}_{i,\sigma} c_{i-1,\sigma}+h.c.)\right]
    \nonumber\\
    &+\sum_{\sigma,i=\mk{2}}^{n_{\rm L}}
    \left[\epsilon^{v}_{i,\sigma} v^{\dag}_{i,\sigma} v_{i,\sigma}
    +V^{v}_{i,\sigma} (v^{\dag}_{i,\sigma}v_{i-1,\sigma}+h.c.) \right]
    , \\
    {\cal H}_{\rm chain} 
    =& 
    \sum_{\sigma,i=n_{\rm L}+1}^{n^{c}_{\sigma}}
    \left[\epsilon^{c}_{i,\sigma} c^{\dag}_{i,\sigma} c_{i,\sigma}
    +V^{c}_{i,\sigma} 
    (c^{\dag}_{i,\sigma} c_{i-1,\sigma}+h.c.) \right]
    \nonumber\\
    &+\sum_{\sigma,i=n_{\rm L}+1}^{n^{v}_{\sigma}}
    \left[\epsilon^{v}_{i,\sigma} v^{\dag}_{i,\sigma}v_{i,\sigma}
    +V^{v}_{i,\sigma} (v^{\dag}_{i,\sigma} v_{i-1,\sigma}+h.c.)\right],
\end{align}
where $\epsilon^{c(v)}_{i,\sigma}, V^{c(v)}_{i,\sigma}$, and  $n^{c(v)}_{i,\sigma}$ 
are the energy levels, hoppings of conduction (valence) sites and the number of these sites, respectively. 
\mk{$V^{{\rm imp},b}$ and $V^{{\rm imp},c}(V^{{\rm imp},v})$ are hoppings between the impurity site and the active bath site and the first conduction (valence) site.}
$b^{\dag}_{\sigma}(b_{\sigma}),c^{\dag}_{i,\sigma}(c_{i,\sigma})$, and $v^{\dag}_{i,\sigma}(v_{i,\sigma})$ are the electron creation (annihilation) operators at the active bath site, $i$-th conduction-bath site, and $i$-th valence-bath site with spin $\sigma$, respectively. 
The number of electrons in the small system,
$(n^{\rm I}_{\uparrow},n^{\rm I}_{\downarrow})$, is automatically determined as
\begin{align}
    n^{\rm I}_{\sigma}
    =n_{\sigma}-n^{v}_{\sigma}+n_{\rm L},
\end{align}
and then we perform the exact diagonalization for ${\cal H}_{\rm I}$ with ($n^{\rm I}_{\uparrow},n^{\rm I}_{\downarrow}$) electrons. From this, the energy of the whole system $(E_{\rm AIM})$ is calculated as
\begin{align}
    E_{\rm AIM}=E_{{\cal H}_{\rm I}} 
    + 
    \sum_{\sigma,i=n_{\rm L}+1}^{n^{v}_{\sigma}} 
    \epsilon^{v}_{i,\sigma},
\end{align}
where $E_{{\cal H}_{\rm I}}$ is the direct result from the ED.
The contribution to the Green function from each sector is computed by the projection method proposed in Ref.~\onlinecite{Lu2019} with $p=2$.
\mk{After obtaining eigenstates of ${\cal H}_{\rm I}$, we construct $n_M$ Krylov vectors ${\cal H}_{\rm I}^j a^{\dag}_{\rm imp} |{\Psi}\rangle (j=1,\cdots,n_M)$ for each eigenstate, $|{\Psi} \rangle$, within ${\cal H}_{\rm I}$ space. From that Krylov subspace, we consider up to two excitations (two electrons, electron-hole, or two holes) into the conduction (valence) chain (${\cal H}_{\rm chain}$) when tridiagonalizing the whole Hamiltonian for calculating the Green function\cite{note_orthonormal}.}
Since the calculation for the Green function of each sector is fully independent of each other, we can just sum-up all the 
sector contributions with the Boltzmann weight $e^{-\beta(E_{\rm AIM}-E_{\rm GS})}$, where $E_{\rm GS}$ is the ground-state energy of the whole system. 

After obtaining the Green function, we complete the DMFT self-consistent loop in a usual way. Since we employ a large number of bath sites to represent a real frequency structure, we can perform the bath-fitting directly at real frequencies with equal weight configurations\cite{PhysRevB.71.045122} to determine the bath parameters in Eq.~\eqref{Eq:bath-ham}.


\begin{figure}[htbp]
        \centering
                \includegraphics[width=\linewidth,angle=0]{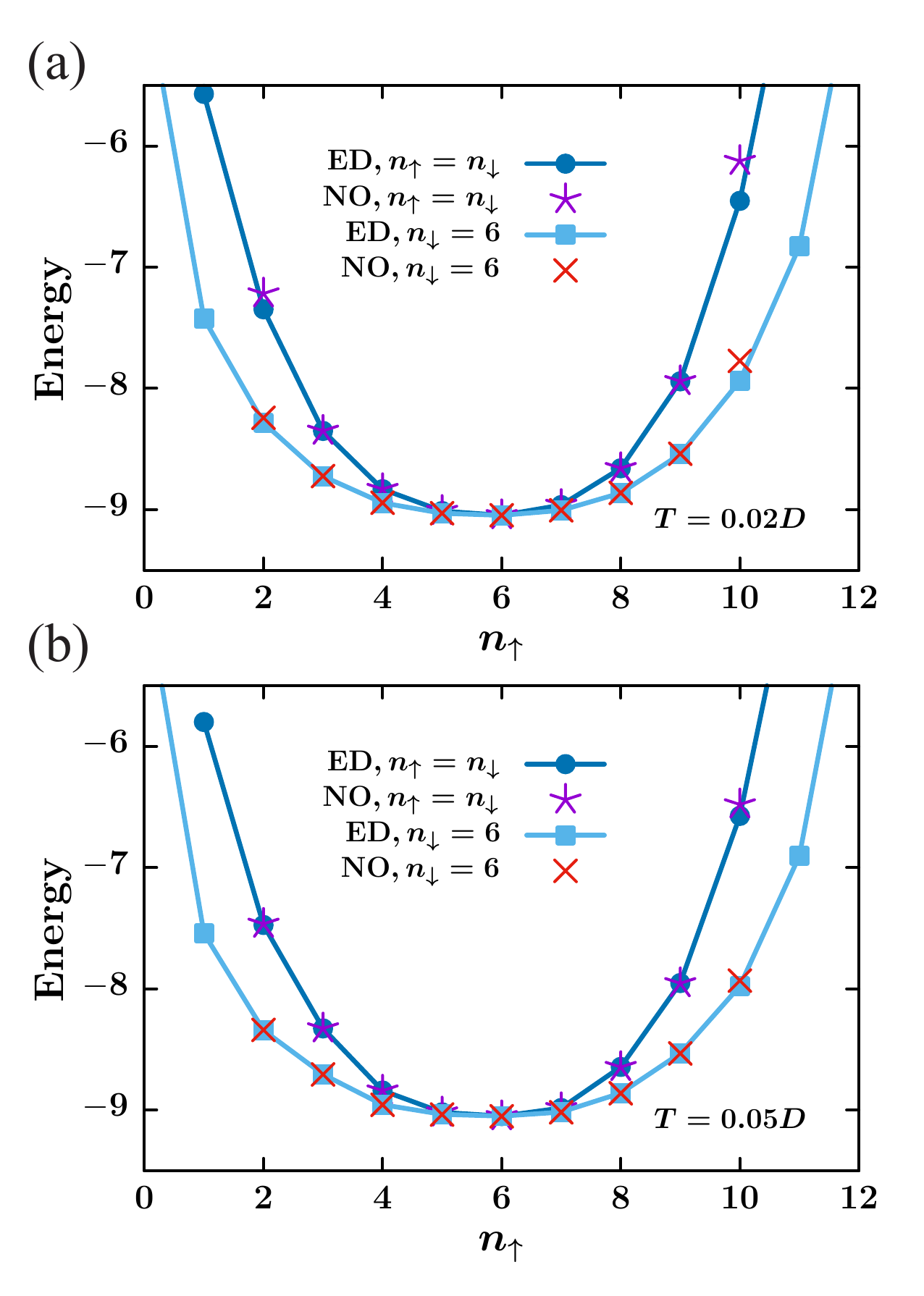}
                \caption{Comparison of the lowest energy of each $(n_{\uparrow},n_{\downarrow})$ sector between the 12-site exact diagonalization (ED) and the 6-site ED with the natural orbital configuration (NO). Two cuts with $n_{\downarrow}=n_{\uparrow}$ and $n_{\downarrow}=6$ are shown for (a) $T=0.02D$ and (b) $T=0.05D$.
        \label{Fig:bench_energy}}
\end{figure}

\section{Results}
\subsection{Benchmarks for a small cluster}
In this section, we start with some benchmarks of finite-temperature results for a small cluster. First, we study the 12-site impurity problem (one impurity site and eleven bath sites) and test the accuracy of the natural orbital solver against the direct ED calculation. For the natural orbital calculation, we diagonalize 6-site ${\cal H}_{\rm I}$ problem (i.e., $n_{\rm L}=2$) with assuming a fixed configuration for the rest 6 bath degrees of freedom.
Namely, we construct conduction- and valence-bath chains with these 6 degrees of freedom and connect them to ${\cal H}_{\rm I}$.
We employ the semicircular density of states $\rho(\epsilon)=2 \sqrt{(1-\omega/D)^2} /\pi D$ and $U=D$, where $D$ is the half of the bandwidth. The electron density is set to \mk{$n=n_{\uparrow}+n_{\downarrow}=0.8$, which makes the system hole-doped but still close to half-filling}.

In Fig.~\ref{Fig:bench_energy},  we first show the lowest energy at each sector, $(n_{\uparrow}, n_{\downarrow})$, obtained with the ED and natural orbital calculation. The parameters of the impurity model are determined from the converged 12-site DMFT-ED calculations at two different temperatures, (a) $T=0.02D$ and (b) $T=0.05D$. From the ED calculations, $(n_{\uparrow},n_{\downarrow})=(6,6)$ is the ground-state sector, so that we show $n_{\uparrow}$ dependence of the lowest energy of each sector for $n_{\downarrow}=6$ (circles) and $n_{\downarrow}=n_{\uparrow}$ (squares). The results show a good agreement in particular at low energies which are important to calculate the Green function accurately. At $n_{\uparrow}=2$ and $10$, we find some deviation from the ED result which suggests that the natural orbital representation becomes worse for the sectors that the length of the chain becomes nearly zero, i.e., $n^{\rm c}_{\sigma}$ or $n^{\rm v}_{\sigma} \sim 0$. In general, such states would have a relatively high energy and their contribution to the Green function is small. Furthermore, such contribution becomes more and more negligible as the total number of bath sites (chain length) increases, for which we will be able to reach a few hundreds in the following. Thus, we can conclude that the natural orbital configuration has an accuracy enough to describe the energy not only for the ground-state sector but also for all energy configurations of the system.

\begin{figure}[htbp]
        \centering
                \includegraphics[width=0.9\linewidth,angle=0]{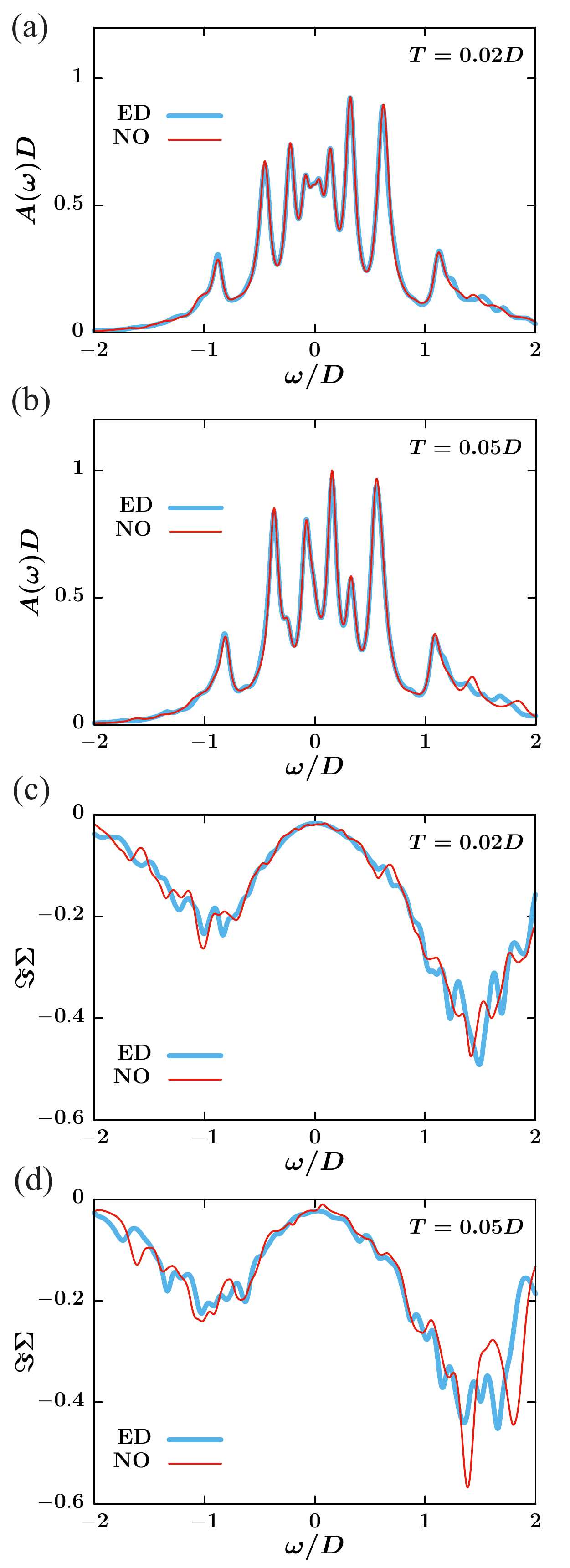}
                \caption{\mk{Comparison of (a-b) the spectrum and (c-d) the imaginary part of the self-energy between the 12-site exact diagonalization (ED) and the 6-site ED with the natural orbital configuration (NO).}
        \label{Fig:bench_spectrum}}
\end{figure}

We also compare the spectrum in Fig.~\ref{Fig:bench_spectrum}\mk{(a,b)} which is obtained from the imaginary part of the Green function as $A(\omega)=-\Im G(\omega)/\pi$\mk{, and the imaginary part of the self-energy $\Im\Sigma(\omega)$ in Fig.~\ref{Fig:bench_spectrum}(c,d)}. Since we are here focusing on the small cluster with a spiky spectrum, we employ a relatively large smearing factor $\gamma=0.05D$ (i.e., $\omega \rightarrow \omega+i\gamma$ when calculating \mk{$G(\omega)$ and $\Sigma(\omega)$}
). The agreement here is also quite good and especially, \mk{spectrums} match perfectly at a low-frequency region. This result demonstrates that the projection method\cite{Lu2019} with $p=2$ (i.e., up to double particle-hole excitations to ${\cal H}_{\rm chain}$) for calculating the Green function works quite well in the temperature region we study.

\subsection{Spectrum for \mk{a} large system}
Now we move to the result with \mk{a} large number of bath sites which can not be addressed by the original ED method. Here, we take 301 bath sites and 10 sites for the ED, i.e., $n_{\rm I}=10$ and $n_{\rm L}=4$. For obtaining the smooth spectra, we here convolute spectra with a Gaussian kernel with the full width at half maximum of $0.04D$ following Ref.~\onlinecite{Lu2019}, and employ a smearing factor of $\gamma=0.02D$.

\begin{figure}[tp]
        \centering
                \includegraphics[width=\linewidth,angle=0]{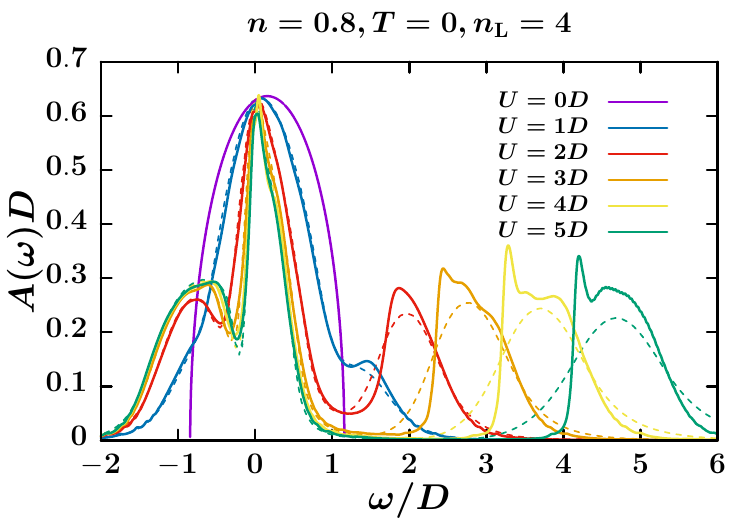}
        \caption{Density of states for the semicircular non-interacting density of states. We take $n_{{\cal H}_{\rm AIM}}=302$ and $n_{{\cal H}_{\rm I}}=10$. Dotted curves are NRG results taken from Ref.~\onlinecite{zitko2015}.
        \label{Fig:DOS-n0.8-semicircular}}
\end{figure}

In Fig.~\ref{Fig:DOS-n0.8-semicircular}, we show the DMFT spectrum at $T=0$ for the semicircular density of states for several interaction strengths $U=1-5D$ at $n=0.8$. 
The previous NRG result \cite{zitko2015} is also shown as dotted curves for comparison. We can see that the results agree very well in a low energy regime and in an overall structure. One difference is in the upper Hubbard band, where our result shows an additional peak at the gap edge. 
While this is not seen in the NRG results, it is known that NRG discards detailed high-frequency structures due to the logarithmic discretization \mk{in energy}. Such a peak was reported at half filling by the recent study using matrix product states\cite{Ganahl2014,Ganahl2015z} and the D-DMRG\cite{PhysRevB.77.075116}, and its origin was further analyzed by Lee {\it et al.} in Refs.~\onlinecite{Lee2017a,Lee2017b}. Our results are consistent with these studies, indicating that the current method works well to obtain the smooth spectrum for calculating real-frequency properties. 


\subsection{Real-frequency properties at finite temperatures}
\begin{figure}[tp]
        \centering
                \includegraphics[width=\linewidth,angle=0]{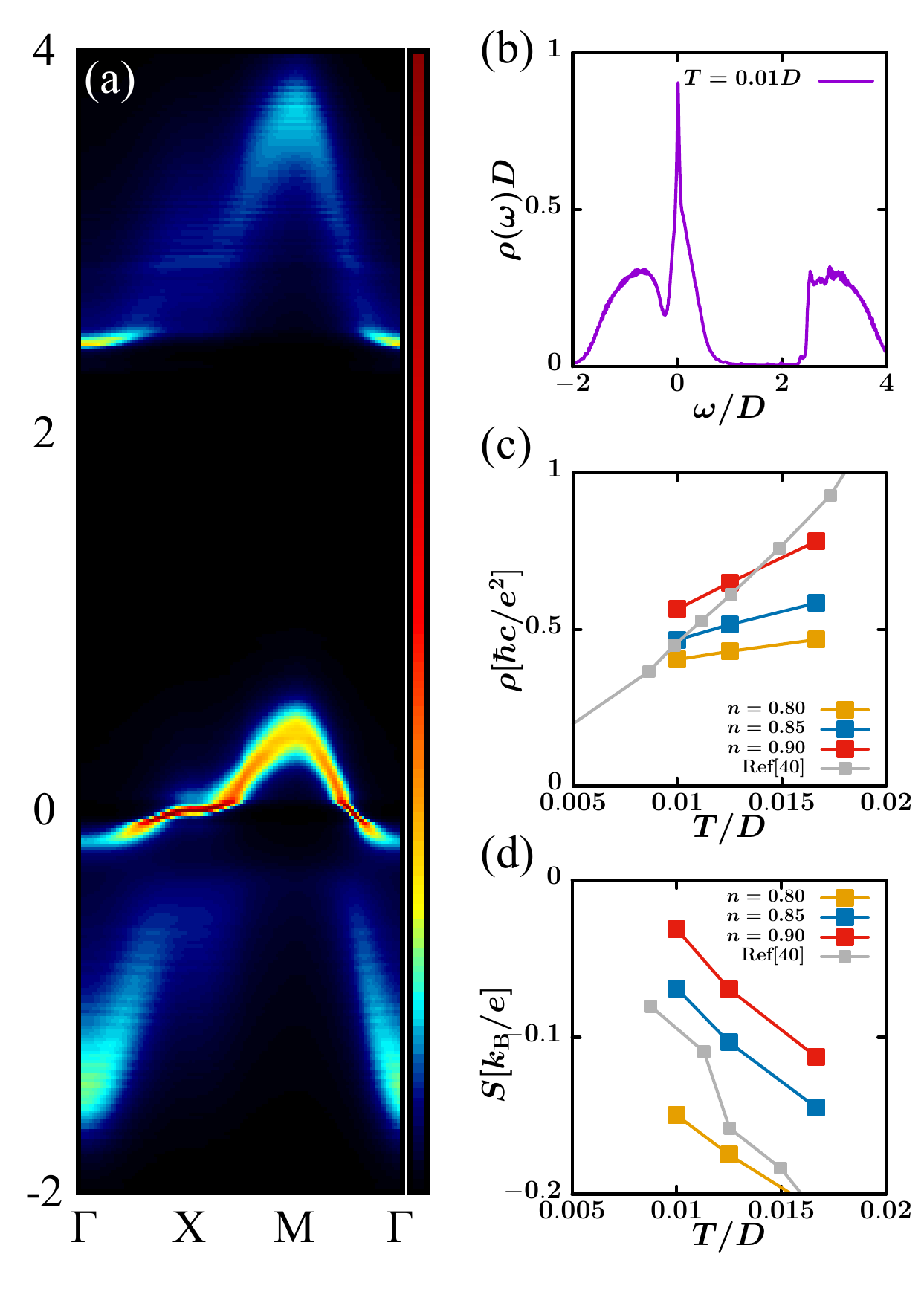}
        \caption{(a) Spectral weight (b) density of states (c) resistivity (d) Seebeck coefficient for the two-dimensional Hubbard model on a square lattice with $U=3.5D$. Units are represented by the universal constants, $\hbar, k_{\rm B}, e$, and the out-of-plane lattice constant c. \mk{Gray lines are the reference CT-QMC+Pad\'{e} results taken from Ref.~\onlinecite{Xu2013}.}
        \label{Fig:2DHubbard}}
\end{figure}

Finally, we show the result of real-frequency properties at finite temperatures. Here, we employ the square lattice Hubbard model with only the nearest neighbor hopping $t=D/4$ and $U=3.5D$. Here, we take 301 bath sites, and 8 sites for the ED calculation, where we employ $\gamma=0.01D$. We sum up the Green function contribution
\mk{from each sector with energy cutoff: $e^{-(E-E_{\rm GS})/T} \geq 0.02$}. 
As shown in Fig.~\ref{Fig:2DHubbard}(a,b), the spectrum obtained with the present method is consistent with the previous CT-QMC+Pad\'{e} results \cite{Xu2013}. We obtain the clear upper Hubbard band structure which is sometimes difficult to obtain through a numerical analytical continuation because it is located at a high energy, i.e., \mk{far} away from the Matsubara axis. The smooth spectrum allows us to calculate the resistivity $\rho$ and the Seebeck coefficient $S$ from\cite{Xu2013}
\begin{align}
    \rho&=\frac{1}{\sigma_1}, 
    S = -\frac{\sigma_2}{\sigma_1},\nonumber\\
    \sigma_i &= 2\pi \int d\omega 
    \left( 
    -\frac{\partial f}{\partial \omega}
    \right) \left(\frac{\omega}{T}\right)^{i-1}
    \sum_{\textbf{k}} A(\omega,\textbf{k})^2v_x(\textbf{k})^2,
\end{align}
where $f$ is the Fermi distribution function and $v_x(\textbf{k}) \equiv 0.5D\sin{k_x}$ is the 
Fermi velocity along the $x$ axis.

In Fig.~\ref{Fig:2DHubbard}(c,d), we show these results at \mk{$n=0.80, 0.85, 0.90$ and} around $T=0.01D$. The temperature dependence is also consistent with the CT-QMC+Pad\'{e} result\mk{\cite{Xu2013} (gray line)}. 
Such a negative slope of the Seebeck coefficient in the low-temperature region of the hole-doped Mott insulator commonly appears in many lattice models, where the correlation plays a crucial role in enhancing its intensity\cite{PhysRevB.87.035126,PhysRevB.88.205141,PhysRevLett.110.086401}. 
\mk{Note that because the results of Ref.~\onlinecite{Xu2013} could be influenced by the error in the Pad\'{e} approximation, a detailed comparison beyond the overall agreement would not be very meaningful.}
Thus, our calculations demonstrate that the natural orbital configuration provides us with an efficient way to address low temperature physics.


\section{Conclusion and outlook}
In conclusion, we have extended the natural orbital impurity solver to finite temperatures by constructing different natural orbitals for each sector in the ED. We have found that the minimum energy of each sector and the Green function are well approximated by this scheme at low but finite temperatures.
We have examined the accuracy of our result for the small cluster against the direct ED calculation and obtained a good agreement. We have also shown that the smooth spectrum can be obtained with a large number of bath sites. Then, we have calculated real-frequency and transport properties for the two-dimensional Hubbard model. Our results demonstrate the usefulness of the natural orbital framework for obtaining these information at finite temperatures.

\mk{While the ED itself is efficient at low temperatures,}
when used as a DMFT impurity solver, it 
\mk{encounters (in the DMFT self-consistent loop) a difficulty in accurately fitting the dynamical mean field with a small number of bath sites particularly at low temperatures.}
The natural orbital scheme removes this problem \mk{by incorporating many bath degrees of freedom} so that we can take the advantage of the original strength of the ED for low temperature calculations.

In this paper, we 
construct distinct natural orbitals for different sectors, and in each sector, we use a fixed set of natural orbitals constructed for describing the accurate ground state. 
Like the finite temperature ED, we consider multiple eigenstates of ${\cal H}_{\rm I}$ in each sector 
on the basis of these
fixed natural orbitals,
while they may not be suitable for \mk{describing} excited states 
of the whole system: ${\cal H}_{\rm I}+{\cal H}_{\rm chain}$.
In this sense, the current treatment may still suffer from the finite size effect which would become \mk{more} important at higher temperatures \mk{(as shown in Appendix)}. Constructing different natural orbitals for each excited state may accelerate the convergence against $n_{\rm L}$ (i.e., reduce the finite size effect).

While the present work focuses on the extension to finite temperatures, the extension to multi-orbitals is also important for the application to real materials. This is another direction of future study. In such applications, the ED has an important advantage in dealing with off-diagonal elements and general-type interactions, which would cause a severe negative sign problem in the CT-QMC methods. \mk{Another promising direction is the extension to the cluster DMFT calculation, which can take spatial fluctuation effect into account with high energy resolution.}

\section*{Appendix}
\begin{figure}[htbp]
        \centering
                \includegraphics[width=\linewidth,angle=0]{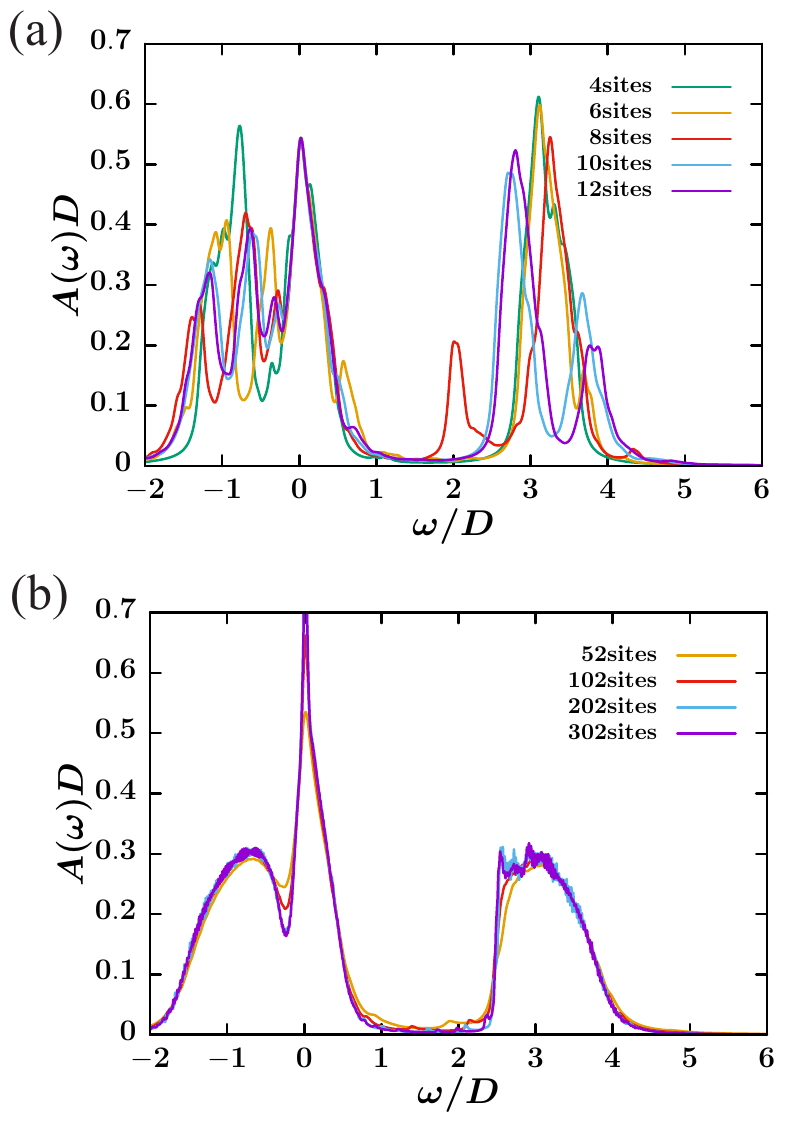}
        \caption{\mk{Total site number dependences of the density of states calculated with (a) the exact diagonalization and (b) the natural orbital framework with $n_{\rm I}=8$.}}
\end{figure}

\mk{In Fig. 6, we show the total site number dependence of the density of states obtained with (a) the exact diagonalization method and (b) the natural orbital framework for the two-dimensional Hubbard model on a square lattice with $U=3.5D, n=0.85, T=0.01D$ (same as Fig.~\ref{Fig:2DHubbard}(b) in the main text). 
Here we set a smearing factor $\gamma=0.05D$ for the exact diagonalization. In Fig.~6(b), we used $\gamma=0.01D$ for 302 sites calculation, and increase $\gamma$ for keeping the ratio of $\gamma$/(the number of bath sites)$^{-1}$ to avoid strong oscillation of the spectrum \cite{Lu2014}. Similar to the original work \cite{Lu2014}, we obtained the almost converged result for the natural orbital framework. On the other hand, the exact diagonalization results strongly depend on the number of the bath sites.}

\begin{figure}[tbp]
        \centering
                \includegraphics[width=\linewidth,angle=0]{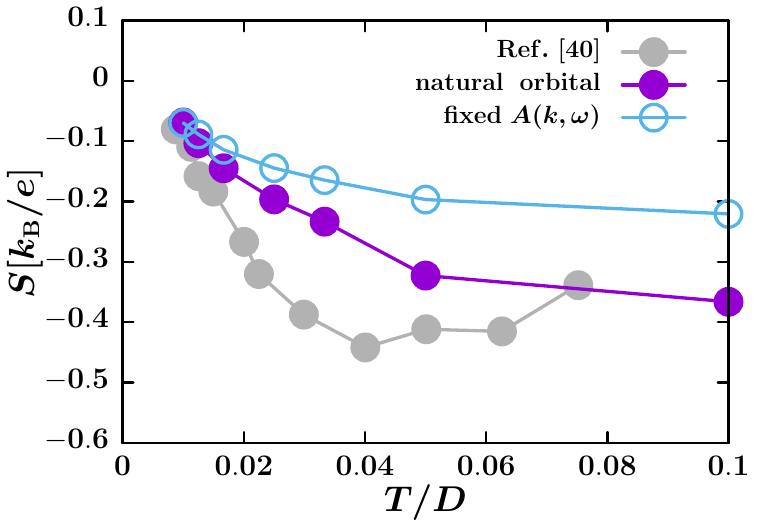}
        \caption{\mk{Wide-range temperature dependence of Seebeck coefficient. Purple dots represent the result of the present method. Blue dots are the result of the fixed spectrum at $\beta D=100$. Gray dots are benchmark (CTQMC+P\'ade) results taken from Ref.~[\onlinecite{Xu2013}].}}
\end{figure}

\mk{Next, we 
examine the accuracy of the current method
for higher temperatures.
Fig.~7 shows the wide-range temperature dependence of the Seebeck coefficient (purple) against the benchmark (gray) taken from Ref.[\onlinecite{Xu2013}]. 
Same as in Sec.~III C, we sum up the Green function contribution with energy cutoff $e^{-(E-E_{\rm GS})/T} \geq 0.02$ except for $T/D=0.1$, where we set $e^{-(E-E_{\rm GS})/T} \geq 0.1$ for quicker convergence.
The results indicate that the agreement for low temperatures is good while it becomes worse for higher temperatures. 
Aside from the ambiguity due to the Pad\'{e} approximation used in Ref.~\onlinecite{Xu2013}, this  would be
because we use the same natural orbital description even for excited states in the same sector as 
mentioned
at the end of the main text. For evaluating the effect of the current finite-temperature extension, we also compared it to the results calculated with a fixed spectrum (result at $T=0.01D$) shown as the blue line in Fig.~7. We can see that the the present method improves the results significantly (blue $\rightarrow$ purple) but not sufficient for high temperatures. There, we would need further development, for example, 
constructing
different natural orbitals for each exited state.
Note that the deviation becomes noticeable for $T/D \gtrsim 0.02$, roughly corresponding to $T \gtrsim 300K$ assuming typical bandwidth $D \sim {\rm (a\;few \;eV)}$. Thus, the current formalism can still be useful to discuss the low-temperature physics of our interest.}

\section*{Acknowledgments}
We thank X. Cao and P. Hansmann for useful discussions. This  work  was  supported  by  a  Grant-in-Aid  for  Scientific  Research  (No. 23H04528, No. 23H03817, No. 21H04437, No. 21H04990 and No. 19H05825), Research Activity Start-up (No. 20K22342), Early-Career Scientists (No. 21K13887), “Program for Promoting Researches on the Supercomputer Fugaku” (Project ID: hp200132, hp210163) from  MEXT, JST-CREST (JPMJCR18T3), and JST-Mirai Program (JPMJMI20A1).

\bibliography{main,naturalorbital}

\begin{thebibliography}{44}%
\makeatletter
\providecommand \@ifxundefined [1]{%
 \@ifx{#1\undefined}
}%
\providecommand \@ifnum [1]{%
 \ifnum #1\expandafter \@firstoftwo
 \else \expandafter \@secondoftwo
 \fi
}%
\providecommand \@ifx [1]{%
 \ifx #1\expandafter \@firstoftwo
 \else \expandafter \@secondoftwo
 \fi
}%
\providecommand \natexlab [1]{#1}%
\providecommand \enquote  [1]{``#1''}%
\providecommand \bibnamefont  [1]{#1}%
\providecommand \bibfnamefont [1]{#1}%
\providecommand \citenamefont [1]{#1}%
\providecommand \href@noop [0]{\@secondoftwo}%
\providecommand \href [0]{\begingroup \@sanitize@url \@href}%
\providecommand \@href[1]{\@@startlink{#1}\@@href}%
\providecommand \@@href[1]{\endgroup#1\@@endlink}%
\providecommand \@sanitize@url [0]{\catcode `\\12\catcode `\$12\catcode
  `\&12\catcode `\#12\catcode `\^12\catcode `\_12\catcode `\%12\relax}%
\providecommand \@@startlink[1]{}%
\providecommand \@@endlink[0]{}%
\providecommand \url  [0]{\begingroup\@sanitize@url \@url }%
\providecommand \@url [1]{\endgroup\@href {#1}{\urlprefix }}%
\providecommand \urlprefix  [0]{URL }%
\providecommand \Eprint [0]{\href }%
\providecommand \doibase [0]{http://dx.doi.org/}%
\providecommand \selectlanguage [0]{\@gobble}%
\providecommand \bibinfo  [0]{\@secondoftwo}%
\providecommand \bibfield  [0]{\@secondoftwo}%
\providecommand \translation [1]{[#1]}%
\providecommand \BibitemOpen [0]{}%
\providecommand \bibitemStop [0]{}%
\providecommand \bibitemNoStop [0]{.\EOS\space}%
\providecommand \EOS [0]{\spacefactor3000\relax}%
\providecommand \BibitemShut  [1]{\csname bibitem#1\endcsname}%
\let\auto@bib@innerbib\@empty
\bibitem [{\citenamefont {Metzner}\ and\ \citenamefont
  {Vollhardt}(1989)}]{Metzner1989}%
  \BibitemOpen
  \bibfield  {author} {\bibinfo {author} {\bibfnamefont {W.}~\bibnamefont
  {Metzner}}\ and\ \bibinfo {author} {\bibfnamefont {D.}~\bibnamefont
  {Vollhardt}},\ }\href {\doibase 10.1103/PhysRevLett.62.324} {\bibfield
  {journal} {\bibinfo  {journal} {Phys. Rev. Lett.}\ }\textbf {\bibinfo
  {volume} {62}},\ \bibinfo {pages} {324} (\bibinfo {year} {1989})}\BibitemShut
  {NoStop}%
\bibitem [{\citenamefont {Georges}\ and\ \citenamefont
  {Krauth}(1992)}]{Georges1992}%
  \BibitemOpen
  \bibfield  {author} {\bibinfo {author} {\bibfnamefont {A.}~\bibnamefont
  {Georges}}\ and\ \bibinfo {author} {\bibfnamefont {W.}~\bibnamefont
  {Krauth}},\ }\href {\doibase 10.1103/PhysRevLett.69.1240} {\bibfield
  {journal} {\bibinfo  {journal} {Phys. Rev. Lett.}\ }\textbf {\bibinfo
  {volume} {69}},\ \bibinfo {pages} {1240} (\bibinfo {year}
  {1992})}\BibitemShut {NoStop}%
\bibitem [{\citenamefont {Georges}\ \emph {et~al.}(1996)\citenamefont
  {Georges}, \citenamefont {Kotliar}, \citenamefont {Krauth},\ and\
  \citenamefont {Rozenberg}}]{Georges1996}%
  \BibitemOpen
  \bibfield  {author} {\bibinfo {author} {\bibfnamefont {A.}~\bibnamefont
  {Georges}}, \bibinfo {author} {\bibfnamefont {G.}~\bibnamefont {Kotliar}},
  \bibinfo {author} {\bibfnamefont {W.}~\bibnamefont {Krauth}}, \ and\ \bibinfo
  {author} {\bibfnamefont {M.~J.}\ \bibnamefont {Rozenberg}},\ }\href {\doibase
  10.1103/RevModPhys.68.13} {\bibfield  {journal} {\bibinfo  {journal} {Rev.
  Mod. Phys.}\ }\textbf {\bibinfo {volume} {68}},\ \bibinfo {pages} {13}
  (\bibinfo {year} {1996})}\BibitemShut {NoStop}%
\bibitem [{\citenamefont {Kotliar}\ \emph {et~al.}(2006)\citenamefont
  {Kotliar}, \citenamefont {Savrasov}, \citenamefont {Haule}, \citenamefont
  {Oudovenko}, \citenamefont {Parcollet},\ and\ \citenamefont
  {Marianetti}}]{Kotliar2006}%
  \BibitemOpen
  \bibfield  {author} {\bibinfo {author} {\bibfnamefont {G.}~\bibnamefont
  {Kotliar}}, \bibinfo {author} {\bibfnamefont {S.~Y.}\ \bibnamefont
  {Savrasov}}, \bibinfo {author} {\bibfnamefont {K.}~\bibnamefont {Haule}},
  \bibinfo {author} {\bibfnamefont {V.~S.}\ \bibnamefont {Oudovenko}}, \bibinfo
  {author} {\bibfnamefont {O.}~\bibnamefont {Parcollet}}, \ and\ \bibinfo
  {author} {\bibfnamefont {C.~A.}\ \bibnamefont {Marianetti}},\ }\href
  {\doibase 10.1103/RevModPhys.78.865} {\bibfield  {journal} {\bibinfo
  {journal} {Rev. Mod. Phys.}\ }\textbf {\bibinfo {volume} {78}},\ \bibinfo
  {pages} {865} (\bibinfo {year} {2006})}\BibitemShut {NoStop}%
\bibitem [{\citenamefont {Held}(2007)}]{Held2007}%
  \BibitemOpen
  \bibfield  {author} {\bibinfo {author} {\bibfnamefont {K.}~\bibnamefont
  {Held}},\ }\href {\doibase 10.1080/00018730701619647} {\bibfield  {journal}
  {\bibinfo  {journal} {Advances in Physics}\ }\textbf {\bibinfo {volume}
  {56}},\ \bibinfo {pages} {829} (\bibinfo {year} {2007})}\BibitemShut
  {NoStop}%
\bibitem [{\citenamefont {Gull}\ \emph {et~al.}(2011)\citenamefont {Gull},
  \citenamefont {Millis}, \citenamefont {Lichtenstein}, \citenamefont
  {Rubtsov}, \citenamefont {Troyer},\ and\ \citenamefont
  {Werner}}]{Gull-RevModPhys.83.349}%
  \BibitemOpen
  \bibfield  {author} {\bibinfo {author} {\bibfnamefont {E.}~\bibnamefont
  {Gull}}, \bibinfo {author} {\bibfnamefont {A.~J.}\ \bibnamefont {Millis}},
  \bibinfo {author} {\bibfnamefont {A.~I.}\ \bibnamefont {Lichtenstein}},
  \bibinfo {author} {\bibfnamefont {A.~N.}\ \bibnamefont {Rubtsov}}, \bibinfo
  {author} {\bibfnamefont {M.}~\bibnamefont {Troyer}}, \ and\ \bibinfo {author}
  {\bibfnamefont {P.}~\bibnamefont {Werner}},\ }\href {\doibase
  10.1103/RevModPhys.83.349} {\bibfield  {journal} {\bibinfo  {journal} {Rev.
  Mod. Phys.}\ }\textbf {\bibinfo {volume} {83}},\ \bibinfo {pages} {349}
  (\bibinfo {year} {2011})}\BibitemShut {NoStop}%
\bibitem [{\citenamefont {Gubernatis}\ \emph {et~al.}(1991)\citenamefont
  {Gubernatis}, \citenamefont {Jarrell}, \citenamefont {Silver},\ and\
  \citenamefont {Sivia}}]{Gubernatis91}%
  \BibitemOpen
  \bibfield  {author} {\bibinfo {author} {\bibfnamefont {J.~E.}\ \bibnamefont
  {Gubernatis}}, \bibinfo {author} {\bibfnamefont {M.}~\bibnamefont {Jarrell}},
  \bibinfo {author} {\bibfnamefont {R.~N.}\ \bibnamefont {Silver}}, \ and\
  \bibinfo {author} {\bibfnamefont {D.~S.}\ \bibnamefont {Sivia}},\ }\href
  {\doibase 10.1103/PhysRevB.44.6011} {\bibfield  {journal} {\bibinfo
  {journal} {Phys. Rev. B}\ }\textbf {\bibinfo {volume} {44}},\ \bibinfo
  {pages} {6011} (\bibinfo {year} {1991})}\BibitemShut {NoStop}%
\bibitem [{\citenamefont {Otsuki}\ \emph {et~al.}(2017)\citenamefont {Otsuki},
  \citenamefont {Ohzeki}, \citenamefont {Shinaoka},\ and\ \citenamefont
  {Yoshimi}}]{Otsuki17}%
  \BibitemOpen
  \bibfield  {author} {\bibinfo {author} {\bibfnamefont {J.}~\bibnamefont
  {Otsuki}}, \bibinfo {author} {\bibfnamefont {M.}~\bibnamefont {Ohzeki}},
  \bibinfo {author} {\bibfnamefont {H.}~\bibnamefont {Shinaoka}}, \ and\
  \bibinfo {author} {\bibfnamefont {K.}~\bibnamefont {Yoshimi}},\ }\href
  {\doibase 10.1103/PhysRevE.95.061302} {\bibfield  {journal} {\bibinfo
  {journal} {Phys. Rev. E}\ }\textbf {\bibinfo {volume} {95}},\ \bibinfo
  {pages} {061302} (\bibinfo {year} {2017})}\BibitemShut {NoStop}%
\bibitem [{\citenamefont {Fei}\ \emph {et~al.}(2021)\citenamefont {Fei},
  \citenamefont {Yeh},\ and\ \citenamefont {Gull}}]{Fei21}%
  \BibitemOpen
  \bibfield  {author} {\bibinfo {author} {\bibfnamefont {J.}~\bibnamefont
  {Fei}}, \bibinfo {author} {\bibfnamefont {C.-N.}\ \bibnamefont {Yeh}}, \ and\
  \bibinfo {author} {\bibfnamefont {E.}~\bibnamefont {Gull}},\ }\href {\doibase
  10.1103/PhysRevLett.126.056402} {\bibfield  {journal} {\bibinfo  {journal}
  {Phys. Rev. Lett.}\ }\textbf {\bibinfo {volume} {126}},\ \bibinfo {pages}
  {056402} (\bibinfo {year} {2021})}\BibitemShut {NoStop}%
\bibitem [{\citenamefont {McMillan}\ and\ \citenamefont
  {Rowell}(1965)}]{MCMillan-Rowell-PhysRevLett.14.108}%
  \BibitemOpen
  \bibfield  {author} {\bibinfo {author} {\bibfnamefont {W.~L.}\ \bibnamefont
  {McMillan}}\ and\ \bibinfo {author} {\bibfnamefont {J.~M.}\ \bibnamefont
  {Rowell}},\ }\href {\doibase 10.1103/PhysRevLett.14.108} {\bibfield
  {journal} {\bibinfo  {journal} {Phys. Rev. Lett.}\ }\textbf {\bibinfo
  {volume} {14}},\ \bibinfo {pages} {108} (\bibinfo {year} {1965})}\BibitemShut
  {NoStop}%
\bibitem [{\citenamefont {Bulla}(1999)}]{Bulla99}%
  \BibitemOpen
  \bibfield  {author} {\bibinfo {author} {\bibfnamefont {R.}~\bibnamefont
  {Bulla}},\ }\href {\doibase 10.1103/PhysRevLett.83.136} {\bibfield  {journal}
  {\bibinfo  {journal} {Phys. Rev. Lett.}\ }\textbf {\bibinfo {volume} {83}},\
  \bibinfo {pages} {136} (\bibinfo {year} {1999})}\BibitemShut {NoStop}%
\bibitem [{\citenamefont {Bulla}\ \emph {et~al.}(2008)\citenamefont {Bulla},
  \citenamefont {Costi},\ and\ \citenamefont
  {Pruschke}}]{Bulla-RevModPhys.80.395}%
  \BibitemOpen
  \bibfield  {author} {\bibinfo {author} {\bibfnamefont {R.}~\bibnamefont
  {Bulla}}, \bibinfo {author} {\bibfnamefont {T.~A.}\ \bibnamefont {Costi}}, \
  and\ \bibinfo {author} {\bibfnamefont {T.}~\bibnamefont {Pruschke}},\ }\href
  {\doibase 10.1103/RevModPhys.80.395} {\bibfield  {journal} {\bibinfo
  {journal} {Rev. Mod. Phys.}\ }\textbf {\bibinfo {volume} {80}},\ \bibinfo
  {pages} {395} (\bibinfo {year} {2008})}\BibitemShut {NoStop}%
\bibitem [{\citenamefont {White}(1992)}]{White92}%
  \BibitemOpen
  \bibfield  {author} {\bibinfo {author} {\bibfnamefont {S.~R.}\ \bibnamefont
  {White}},\ }\href {\doibase 10.1103/PhysRevLett.69.2863} {\bibfield
  {journal} {\bibinfo  {journal} {Phys. Rev. Lett.}\ }\textbf {\bibinfo
  {volume} {69}},\ \bibinfo {pages} {2863} (\bibinfo {year}
  {1992})}\BibitemShut {NoStop}%
\bibitem [{\citenamefont {Hallberg}(2006)}]{Hallberg06}%
  \BibitemOpen
  \bibfield  {author} {\bibinfo {author} {\bibfnamefont {K.~A.}\ \bibnamefont
  {Hallberg}},\ }\href {\doibase 10.1080/00018730600766432} {\bibfield
  {journal} {\bibinfo  {journal} {Advances in Physics}\ }\textbf {\bibinfo
  {volume} {55}},\ \bibinfo {pages} {477} (\bibinfo {year} {2006})},\ \Eprint
  {http://arxiv.org/abs/https://doi.org/10.1080/00018730600766432}
  {https://doi.org/10.1080/00018730600766432} \BibitemShut {NoStop}%
\bibitem [{\citenamefont {Schollwöck}(2011)}]{Schollwock11}%
  \BibitemOpen
  \bibfield  {author} {\bibinfo {author} {\bibfnamefont {U.}~\bibnamefont
  {Schollwöck}},\ }\href {\doibase https://doi.org/10.1016/j.aop.2010.09.012}
  {\bibfield  {journal} {\bibinfo  {journal} {Annals of Physics}\ }\textbf
  {\bibinfo {volume} {326}},\ \bibinfo {pages} {96} (\bibinfo {year} {2011})},\
  \bibinfo {note} {january 2011 Special Issue}\BibitemShut {NoStop}%
\bibitem [{\citenamefont {Pruschke}\ and\ \citenamefont
  {Bulla}(2005)}]{PRuschke-Bulla-EPJ.44.217}%
  \BibitemOpen
  \bibfield  {author} {\bibinfo {author} {\bibfnamefont {T.}~\bibnamefont
  {Pruschke}}\ and\ \bibinfo {author} {\bibfnamefont {R.}~\bibnamefont
  {Bulla}},\ }\href@noop {} {\bibfield  {journal} {\bibinfo  {journal} {The
  European Physical Journal B - Condensed Matter and Complex Systems}\ }\textbf
  {\bibinfo {volume} {44}},\ \bibinfo {pages} {217} (\bibinfo {year}
  {2005})}\BibitemShut {NoStop}%
\bibitem [{\citenamefont {Peters}\ and\ \citenamefont
  {Pruschke}(2010)}]{Peters-Pruschke-PhysRevB.81.035112}%
  \BibitemOpen
  \bibfield  {author} {\bibinfo {author} {\bibfnamefont {R.}~\bibnamefont
  {Peters}}\ and\ \bibinfo {author} {\bibfnamefont {T.}~\bibnamefont
  {Pruschke}},\ }\href {\doibase 10.1103/PhysRevB.81.035112} {\bibfield
  {journal} {\bibinfo  {journal} {Phys. Rev. B}\ }\textbf {\bibinfo {volume}
  {81}},\ \bibinfo {pages} {035112} (\bibinfo {year} {2010})}\BibitemShut
  {NoStop}%
\bibitem [{\citenamefont {Stadler}\ \emph {et~al.}(2015)\citenamefont
  {Stadler}, \citenamefont {Yin}, \citenamefont {von Delft}, \citenamefont
  {Kotliar},\ and\ \citenamefont
  {Weichselbaum}}]{Stadler-PhysRevLett.115.136401}%
  \BibitemOpen
  \bibfield  {author} {\bibinfo {author} {\bibfnamefont {K.~M.}\ \bibnamefont
  {Stadler}}, \bibinfo {author} {\bibfnamefont {Z.~P.}\ \bibnamefont {Yin}},
  \bibinfo {author} {\bibfnamefont {J.}~\bibnamefont {von Delft}}, \bibinfo
  {author} {\bibfnamefont {G.}~\bibnamefont {Kotliar}}, \ and\ \bibinfo
  {author} {\bibfnamefont {A.}~\bibnamefont {Weichselbaum}},\ }\href {\doibase
  10.1103/PhysRevLett.115.136401} {\bibfield  {journal} {\bibinfo  {journal}
  {Phys. Rev. Lett.}\ }\textbf {\bibinfo {volume} {115}},\ \bibinfo {pages}
  {136401} (\bibinfo {year} {2015})}\BibitemShut {NoStop}%
\bibitem [{\citenamefont {Kugler}\ \emph {et~al.}(2020)\citenamefont {Kugler},
  \citenamefont {Zingl}, \citenamefont {Strand}, \citenamefont {Lee},
  \citenamefont {von Delft},\ and\ \citenamefont {Georges}}]{Kugler2020}%
  \BibitemOpen
  \bibfield  {author} {\bibinfo {author} {\bibfnamefont {F.~B.}\ \bibnamefont
  {Kugler}}, \bibinfo {author} {\bibfnamefont {M.}~\bibnamefont {Zingl}},
  \bibinfo {author} {\bibfnamefont {H.~U.~R.}\ \bibnamefont {Strand}}, \bibinfo
  {author} {\bibfnamefont {S.-S.~B.}\ \bibnamefont {Lee}}, \bibinfo {author}
  {\bibfnamefont {J.}~\bibnamefont {von Delft}}, \ and\ \bibinfo {author}
  {\bibfnamefont {A.}~\bibnamefont {Georges}},\ }\href {\doibase
  10.1103/PhysRevLett.124.016401} {\bibfield  {journal} {\bibinfo  {journal}
  {Phys. Rev. Lett.}\ }\textbf {\bibinfo {volume} {124}},\ \bibinfo {pages}
  {016401} (\bibinfo {year} {2020})}\BibitemShut {NoStop}%
\bibitem [{\citenamefont {Bauernfeind}\ \emph {et~al.}(2017)\citenamefont
  {Bauernfeind}, \citenamefont {Zingl}, \citenamefont {Triebl}, \citenamefont
  {Aichhorn},\ and\ \citenamefont {Evertz}}]{PhysRevX.7.031013}%
  \BibitemOpen
  \bibfield  {author} {\bibinfo {author} {\bibfnamefont {D.}~\bibnamefont
  {Bauernfeind}}, \bibinfo {author} {\bibfnamefont {M.}~\bibnamefont {Zingl}},
  \bibinfo {author} {\bibfnamefont {R.}~\bibnamefont {Triebl}}, \bibinfo
  {author} {\bibfnamefont {M.}~\bibnamefont {Aichhorn}}, \ and\ \bibinfo
  {author} {\bibfnamefont {H.~G.}\ \bibnamefont {Evertz}},\ }\href {\doibase
  10.1103/PhysRevX.7.031013} {\bibfield  {journal} {\bibinfo  {journal} {Phys.
  Rev. X}\ }\textbf {\bibinfo {volume} {7}},\ \bibinfo {pages} {031013}
  (\bibinfo {year} {2017})}\BibitemShut {NoStop}%
\bibitem [{\citenamefont {Caffarel}\ and\ \citenamefont
  {Krauth}(1994)}]{Caffarel-Krauth-PhysRevLett.72.1545}%
  \BibitemOpen
  \bibfield  {author} {\bibinfo {author} {\bibfnamefont {M.}~\bibnamefont
  {Caffarel}}\ and\ \bibinfo {author} {\bibfnamefont {W.}~\bibnamefont
  {Krauth}},\ }\href {\doibase 10.1103/PhysRevLett.72.1545} {\bibfield
  {journal} {\bibinfo  {journal} {Phys. Rev. Lett.}\ }\textbf {\bibinfo
  {volume} {72}},\ \bibinfo {pages} {1545} (\bibinfo {year}
  {1994})}\BibitemShut {NoStop}%
\bibitem [{\citenamefont {Lu}\ \emph {et~al.}(2014)\citenamefont {Lu},
  \citenamefont {H\"oppner}, \citenamefont {Gunnarsson},\ and\ \citenamefont
  {Haverkort}}]{Lu2014}%
  \BibitemOpen
  \bibfield  {author} {\bibinfo {author} {\bibfnamefont {Y.}~\bibnamefont
  {Lu}}, \bibinfo {author} {\bibfnamefont {M.}~\bibnamefont {H\"oppner}},
  \bibinfo {author} {\bibfnamefont {O.}~\bibnamefont {Gunnarsson}}, \ and\
  \bibinfo {author} {\bibfnamefont {M.~W.}\ \bibnamefont {Haverkort}},\ }\href
  {\doibase 10.1103/PhysRevB.90.085102} {\bibfield  {journal} {\bibinfo
  {journal} {Phys. Rev. B}\ }\textbf {\bibinfo {volume} {90}},\ \bibinfo
  {pages} {085102} (\bibinfo {year} {2014})}\BibitemShut {NoStop}%
\bibitem [{\citenamefont {Lu}\ \emph {et~al.}(2019)\citenamefont {Lu},
  \citenamefont {Cao}, \citenamefont {Hansmann},\ and\ \citenamefont
  {Haverkort}}]{Lu2019}%
  \BibitemOpen
  \bibfield  {author} {\bibinfo {author} {\bibfnamefont {Y.}~\bibnamefont
  {Lu}}, \bibinfo {author} {\bibfnamefont {X.}~\bibnamefont {Cao}}, \bibinfo
  {author} {\bibfnamefont {P.}~\bibnamefont {Hansmann}}, \ and\ \bibinfo
  {author} {\bibfnamefont {M.~W.}\ \bibnamefont {Haverkort}},\ }\href {\doibase
  10.1103/PhysRevB.100.115134} {\bibfield  {journal} {\bibinfo  {journal}
  {Phys. Rev. B}\ }\textbf {\bibinfo {volume} {100}},\ \bibinfo {pages}
  {115134} (\bibinfo {year} {2019})}\BibitemShut {NoStop}%
\bibitem [{\citenamefont {Zgid}\ \emph {et~al.}(2012)\citenamefont {Zgid},
  \citenamefont {Gull},\ and\ \citenamefont {Chan}}]{Zgid12}%
  \BibitemOpen
  \bibfield  {author} {\bibinfo {author} {\bibfnamefont {D.}~\bibnamefont
  {Zgid}}, \bibinfo {author} {\bibfnamefont {E.}~\bibnamefont {Gull}}, \ and\
  \bibinfo {author} {\bibfnamefont {G.~K.-L.}\ \bibnamefont {Chan}},\ }\href
  {\doibase 10.1103/PhysRevB.86.165128} {\bibfield  {journal} {\bibinfo
  {journal} {Phys. Rev. B}\ }\textbf {\bibinfo {volume} {86}},\ \bibinfo
  {pages} {165128} (\bibinfo {year} {2012})}\BibitemShut {NoStop}%
\bibitem [{\citenamefont {Lin}\ and\ \citenamefont {Demkov}(2013)}]{Lin13}%
  \BibitemOpen
  \bibfield  {author} {\bibinfo {author} {\bibfnamefont {C.}~\bibnamefont
  {Lin}}\ and\ \bibinfo {author} {\bibfnamefont {A.~A.}\ \bibnamefont
  {Demkov}},\ }\href {\doibase 10.1103/PhysRevB.88.035123} {\bibfield
  {journal} {\bibinfo  {journal} {Phys. Rev. B}\ }\textbf {\bibinfo {volume}
  {88}},\ \bibinfo {pages} {035123} (\bibinfo {year} {2013})}\BibitemShut
  {NoStop}%
\bibitem [{\citenamefont {Go}\ and\ \citenamefont {Millis}(2015)}]{Go15}%
  \BibitemOpen
  \bibfield  {author} {\bibinfo {author} {\bibfnamefont {A.}~\bibnamefont
  {Go}}\ and\ \bibinfo {author} {\bibfnamefont {A.~J.}\ \bibnamefont
  {Millis}},\ }\href {\doibase 10.1103/PhysRevLett.114.016402} {\bibfield
  {journal} {\bibinfo  {journal} {Phys. Rev. Lett.}\ }\textbf {\bibinfo
  {volume} {114}},\ \bibinfo {pages} {016402} (\bibinfo {year}
  {2015})}\BibitemShut {NoStop}%
\bibitem [{\citenamefont {Go}\ and\ \citenamefont {Millis}(2017)}]{Go17}%
  \BibitemOpen
  \bibfield  {author} {\bibinfo {author} {\bibfnamefont {A.}~\bibnamefont
  {Go}}\ and\ \bibinfo {author} {\bibfnamefont {A.~J.}\ \bibnamefont
  {Millis}},\ }\href {\doibase 10.1103/PhysRevB.96.085139} {\bibfield
  {journal} {\bibinfo  {journal} {Phys. Rev. B}\ }\textbf {\bibinfo {volume}
  {96}},\ \bibinfo {pages} {085139} (\bibinfo {year} {2017})}\BibitemShut
  {NoStop}%
\bibitem [{\citenamefont {Cao}\ \emph {et~al.}(2021)\citenamefont {Cao},
  \citenamefont {Lu}, \citenamefont {Hansmann},\ and\ \citenamefont
  {Haverkort}}]{cao2021tree}%
  \BibitemOpen
  \bibfield  {author} {\bibinfo {author} {\bibfnamefont {X.}~\bibnamefont
  {Cao}}, \bibinfo {author} {\bibfnamefont {Y.}~\bibnamefont {Lu}}, \bibinfo
  {author} {\bibfnamefont {P.}~\bibnamefont {Hansmann}}, \ and\ \bibinfo
  {author} {\bibfnamefont {M.~W.}\ \bibnamefont {Haverkort}},\ }\href {\doibase
  10.1103/PhysRevB.104.115119} {\bibfield  {journal} {\bibinfo  {journal}
  {Phys. Rev. B}\ }\textbf {\bibinfo {volume} {104}},\ \bibinfo {pages}
  {115119} (\bibinfo {year} {2021})}\BibitemShut {NoStop}%
\bibitem [{\citenamefont {Liebsch}\ and\ \citenamefont
  {Ishida}(2012)}]{Liebsch12}%
  \BibitemOpen
  \bibfield  {author} {\bibinfo {author} {\bibfnamefont {A.}~\bibnamefont
  {Liebsch}}\ and\ \bibinfo {author} {\bibfnamefont {H.}~\bibnamefont
  {Ishida}},\ }\href {http://stacks.iop.org/0953-8984/24/i=5/a=053201}
  {\bibfield  {journal} {\bibinfo  {journal} {J. Phys.: Condens. Matter}\
  }\textbf {\bibinfo {volume} {24}},\ \bibinfo {pages} {053201} (\bibinfo
  {year} {2012})}\BibitemShut {NoStop}%
\bibitem [{\citenamefont {Capone}\ \emph {et~al.}(2007)\citenamefont {Capone},
  \citenamefont {de' Medici},\ and\ \citenamefont {Georges}}]{Capone07}%
  \BibitemOpen
  \bibfield  {author} {\bibinfo {author} {\bibfnamefont {M.}~\bibnamefont
  {Capone}}, \bibinfo {author} {\bibfnamefont {L.}~\bibnamefont {de' Medici}},
  \ and\ \bibinfo {author} {\bibfnamefont {A.}~\bibnamefont {Georges}},\ }\href
  {\doibase 10.1103/PhysRevB.76.245116} {\bibfield  {journal} {\bibinfo
  {journal} {Phys. Rev. B}\ }\textbf {\bibinfo {volume} {76}},\ \bibinfo
  {pages} {245116} (\bibinfo {year} {2007})}\BibitemShut {NoStop}%
\bibitem [{\citenamefont {Kohn}\ and\ \citenamefont
  {Santoro}(2021)}]{Kohn2021}%
  \BibitemOpen
  \bibfield  {author} {\bibinfo {author} {\bibfnamefont {L.}~\bibnamefont
  {Kohn}}\ and\ \bibinfo {author} {\bibfnamefont {G.~E.}\ \bibnamefont
  {Santoro}},\ }\href {\doibase 10.1103/PhysRevB.104.014303} {\bibfield
  {journal} {\bibinfo  {journal} {Phys. Rev. B}\ }\textbf {\bibinfo {volume}
  {104}},\ \bibinfo {pages} {014303} (\bibinfo {year} {2021})}\BibitemShut
  {NoStop}%
\bibitem [{\citenamefont {Parragh}\ \emph {et~al.}(2012)\citenamefont
  {Parragh}, \citenamefont {Toschi}, \citenamefont {Held},\ and\ \citenamefont
  {Sangiovanni}}]{Parragh2012}%
  \BibitemOpen
  \bibfield  {author} {\bibinfo {author} {\bibfnamefont {N.}~\bibnamefont
  {Parragh}}, \bibinfo {author} {\bibfnamefont {A.}~\bibnamefont {Toschi}},
  \bibinfo {author} {\bibfnamefont {K.}~\bibnamefont {Held}}, \ and\ \bibinfo
  {author} {\bibfnamefont {G.}~\bibnamefont {Sangiovanni}},\ }\href {\doibase
  10.1103/PhysRevB.86.155158} {\bibfield  {journal} {\bibinfo  {journal} {Phys.
  Rev. B}\ }\textbf {\bibinfo {volume} {86}},\ \bibinfo {pages} {155158}
  (\bibinfo {year} {2012})}\BibitemShut {NoStop}%
\bibitem [{not()}]{note_orthonormal}%
  \BibitemOpen
  \href@noop {} {}\bibinfo {note} {Since the extended states after simply
  applying creation/annihilation operators are not orthonormal, we solve the
  generalized eigenvalue problem to bring them into an orthonormal, the same as
  in Ref.~[\onlinecite{{Lu2019}}].}\BibitemShut {Stop}%
\bibitem [{\citenamefont {Bulla}\ \emph {et~al.}(2005)\citenamefont {Bulla},
  \citenamefont {Lee}, \citenamefont {Tong},\ and\ \citenamefont
  {Vojta}}]{PhysRevB.71.045122}%
  \BibitemOpen
  \bibfield  {author} {\bibinfo {author} {\bibfnamefont {R.}~\bibnamefont
  {Bulla}}, \bibinfo {author} {\bibfnamefont {H.-J.}\ \bibnamefont {Lee}},
  \bibinfo {author} {\bibfnamefont {N.-H.}\ \bibnamefont {Tong}}, \ and\
  \bibinfo {author} {\bibfnamefont {M.}~\bibnamefont {Vojta}},\ }\href
  {\doibase 10.1103/PhysRevB.71.045122} {\bibfield  {journal} {\bibinfo
  {journal} {Phys. Rev. B}\ }\textbf {\bibinfo {volume} {71}},\ \bibinfo
  {pages} {045122} (\bibinfo {year} {2005})}\BibitemShut {NoStop}%
\bibitem [{\citenamefont {\ifmmode~\check{Z}\else \v{Z}\fi{}itko}\ \emph
  {et~al.}(2015)\citenamefont {\ifmmode~\check{Z}\else \v{Z}\fi{}itko},
  \citenamefont {Osolin},\ and\ \citenamefont {Jegli\ifmmode~\check{c}\else
  \v{c}\fi{}}}]{zitko2015}%
  \BibitemOpen
  \bibfield  {author} {\bibinfo {author} {\bibfnamefont {R.}~\bibnamefont
  {\ifmmode~\check{Z}\else \v{Z}\fi{}itko}}, \bibinfo {author} {\bibfnamefont
  {i.~c.~v.}\ \bibnamefont {Osolin}}, \ and\ \bibinfo {author} {\bibfnamefont
  {P.}~\bibnamefont {Jegli\ifmmode~\check{c}\else \v{c}\fi{}}},\ }\href
  {\doibase 10.1103/PhysRevB.91.155111} {\bibfield  {journal} {\bibinfo
  {journal} {Phys. Rev. B}\ }\textbf {\bibinfo {volume} {91}},\ \bibinfo
  {pages} {155111} (\bibinfo {year} {2015})}\BibitemShut {NoStop}%
\bibitem [{\citenamefont {Ganahl}\ \emph {et~al.}(2014)\citenamefont {Ganahl},
  \citenamefont {Thunstr\"om}, \citenamefont {Verstraete}, \citenamefont
  {Held},\ and\ \citenamefont {Evertz}}]{Ganahl2014}%
  \BibitemOpen
  \bibfield  {author} {\bibinfo {author} {\bibfnamefont {M.}~\bibnamefont
  {Ganahl}}, \bibinfo {author} {\bibfnamefont {P.}~\bibnamefont {Thunstr\"om}},
  \bibinfo {author} {\bibfnamefont {F.}~\bibnamefont {Verstraete}}, \bibinfo
  {author} {\bibfnamefont {K.}~\bibnamefont {Held}}, \ and\ \bibinfo {author}
  {\bibfnamefont {H.~G.}\ \bibnamefont {Evertz}},\ }\href {\doibase
  10.1103/PhysRevB.90.045144} {\bibfield  {journal} {\bibinfo  {journal} {Phys.
  Rev. B}\ }\textbf {\bibinfo {volume} {90}},\ \bibinfo {pages} {045144}
  (\bibinfo {year} {2014})}\BibitemShut {NoStop}%
\bibitem [{\citenamefont {Ganahl}\ \emph {et~al.}(2015)\citenamefont {Ganahl},
  \citenamefont {Aichhorn}, \citenamefont {Evertz}, \citenamefont
  {Thunstr\"om}, \citenamefont {Held},\ and\ \citenamefont
  {Verstraete}}]{Ganahl2015z}%
  \BibitemOpen
  \bibfield  {author} {\bibinfo {author} {\bibfnamefont {M.}~\bibnamefont
  {Ganahl}}, \bibinfo {author} {\bibfnamefont {M.}~\bibnamefont {Aichhorn}},
  \bibinfo {author} {\bibfnamefont {H.~G.}\ \bibnamefont {Evertz}}, \bibinfo
  {author} {\bibfnamefont {P.}~\bibnamefont {Thunstr\"om}}, \bibinfo {author}
  {\bibfnamefont {K.}~\bibnamefont {Held}}, \ and\ \bibinfo {author}
  {\bibfnamefont {F.}~\bibnamefont {Verstraete}},\ }\href {\doibase
  10.1103/PhysRevB.92.155132} {\bibfield  {journal} {\bibinfo  {journal} {Phys.
  Rev. B}\ }\textbf {\bibinfo {volume} {92}},\ \bibinfo {pages} {155132}
  (\bibinfo {year} {2015})}\BibitemShut {NoStop}%
\bibitem [{\citenamefont {Karski}\ \emph {et~al.}(2008)\citenamefont {Karski},
  \citenamefont {Raas},\ and\ \citenamefont {Uhrig}}]{PhysRevB.77.075116}%
  \BibitemOpen
  \bibfield  {author} {\bibinfo {author} {\bibfnamefont {M.}~\bibnamefont
  {Karski}}, \bibinfo {author} {\bibfnamefont {C.}~\bibnamefont {Raas}}, \ and\
  \bibinfo {author} {\bibfnamefont {G.~S.}\ \bibnamefont {Uhrig}},\ }\href
  {\doibase 10.1103/PhysRevB.77.075116} {\bibfield  {journal} {\bibinfo
  {journal} {Phys. Rev. B}\ }\textbf {\bibinfo {volume} {77}},\ \bibinfo
  {pages} {075116} (\bibinfo {year} {2008})}\BibitemShut {NoStop}%
\bibitem [{\citenamefont {Lee}\ \emph {et~al.}(2017{\natexlab{a}})\citenamefont
  {Lee}, \citenamefont {von Delft},\ and\ \citenamefont
  {Weichselbaum}}]{Lee2017a}%
  \BibitemOpen
  \bibfield  {author} {\bibinfo {author} {\bibfnamefont {S.-S.~B.}\
  \bibnamefont {Lee}}, \bibinfo {author} {\bibfnamefont {J.}~\bibnamefont {von
  Delft}}, \ and\ \bibinfo {author} {\bibfnamefont {A.}~\bibnamefont
  {Weichselbaum}},\ }\href {\doibase 10.1103/PhysRevLett.119.236402} {\bibfield
   {journal} {\bibinfo  {journal} {Phys. Rev. Lett.}\ }\textbf {\bibinfo
  {volume} {119}},\ \bibinfo {pages} {236402} (\bibinfo {year}
  {2017}{\natexlab{a}})}\BibitemShut {NoStop}%
\bibitem [{\citenamefont {Lee}\ \emph {et~al.}(2017{\natexlab{b}})\citenamefont
  {Lee}, \citenamefont {von Delft},\ and\ \citenamefont
  {Weichselbaum}}]{Lee2017b}%
  \BibitemOpen
  \bibfield  {author} {\bibinfo {author} {\bibfnamefont {S.-S.~B.}\
  \bibnamefont {Lee}}, \bibinfo {author} {\bibfnamefont {J.}~\bibnamefont {von
  Delft}}, \ and\ \bibinfo {author} {\bibfnamefont {A.}~\bibnamefont
  {Weichselbaum}},\ }\href {\doibase 10.1103/PhysRevB.96.245106} {\bibfield
  {journal} {\bibinfo  {journal} {Phys. Rev. B}\ }\textbf {\bibinfo {volume}
  {96}},\ \bibinfo {pages} {245106} (\bibinfo {year}
  {2017}{\natexlab{b}})}\BibitemShut {NoStop}%
\bibitem [{\citenamefont {Xu}\ \emph {et~al.}(2013)\citenamefont {Xu},
  \citenamefont {Haule},\ and\ \citenamefont {Kotliar}}]{Xu2013}%
  \BibitemOpen
  \bibfield  {author} {\bibinfo {author} {\bibfnamefont {W.}~\bibnamefont
  {Xu}}, \bibinfo {author} {\bibfnamefont {K.}~\bibnamefont {Haule}}, \ and\
  \bibinfo {author} {\bibfnamefont {G.}~\bibnamefont {Kotliar}},\ }\href
  {\doibase 10.1103/PhysRevLett.111.036401} {\bibfield  {journal} {\bibinfo
  {journal} {Phys. Rev. Lett.}\ }\textbf {\bibinfo {volume} {111}},\ \bibinfo
  {pages} {036401} (\bibinfo {year} {2013})}\BibitemShut {NoStop}%
\bibitem [{\citenamefont {Arsenault}\ \emph {et~al.}(2013)\citenamefont
  {Arsenault}, \citenamefont {Shastry}, \citenamefont {S\'emon},\ and\
  \citenamefont {Tremblay}}]{PhysRevB.87.035126}%
  \BibitemOpen
  \bibfield  {author} {\bibinfo {author} {\bibfnamefont {L.-F. m.~c.}\
  \bibnamefont {Arsenault}}, \bibinfo {author} {\bibfnamefont {B.~S.}\
  \bibnamefont {Shastry}}, \bibinfo {author} {\bibfnamefont {P.}~\bibnamefont
  {S\'emon}}, \ and\ \bibinfo {author} {\bibfnamefont {A.-M.~S.}\ \bibnamefont
  {Tremblay}},\ }\href {\doibase 10.1103/PhysRevB.87.035126} {\bibfield
  {journal} {\bibinfo  {journal} {Phys. Rev. B}\ }\textbf {\bibinfo {volume}
  {87}},\ \bibinfo {pages} {035126} (\bibinfo {year} {2013})}\BibitemShut
  {NoStop}%
\bibitem [{\citenamefont {Kargarian}\ and\ \citenamefont
  {Fiete}(2013)}]{PhysRevB.88.205141}%
  \BibitemOpen
  \bibfield  {author} {\bibinfo {author} {\bibfnamefont {M.}~\bibnamefont
  {Kargarian}}\ and\ \bibinfo {author} {\bibfnamefont {G.~A.}\ \bibnamefont
  {Fiete}},\ }\href {\doibase 10.1103/PhysRevB.88.205141} {\bibfield  {journal}
  {\bibinfo  {journal} {Phys. Rev. B}\ }\textbf {\bibinfo {volume} {88}},\
  \bibinfo {pages} {205141} (\bibinfo {year} {2013})}\BibitemShut {NoStop}%
\bibitem [{\citenamefont {Deng}\ \emph {et~al.}(2013)\citenamefont {Deng},
  \citenamefont {Mravlje}, \citenamefont {\ifmmode~\check{Z}\else
  \v{Z}\fi{}itko}, \citenamefont {Ferrero}, \citenamefont {Kotliar},\ and\
  \citenamefont {Georges}}]{PhysRevLett.110.086401}%
  \BibitemOpen
  \bibfield  {author} {\bibinfo {author} {\bibfnamefont {X.}~\bibnamefont
  {Deng}}, \bibinfo {author} {\bibfnamefont {J.}~\bibnamefont {Mravlje}},
  \bibinfo {author} {\bibfnamefont {R.}~\bibnamefont {\ifmmode~\check{Z}\else
  \v{Z}\fi{}itko}}, \bibinfo {author} {\bibfnamefont {M.}~\bibnamefont
  {Ferrero}}, \bibinfo {author} {\bibfnamefont {G.}~\bibnamefont {Kotliar}}, \
  and\ \bibinfo {author} {\bibfnamefont {A.}~\bibnamefont {Georges}},\ }\href
  {\doibase 10.1103/PhysRevLett.110.086401} {\bibfield  {journal} {\bibinfo
  {journal} {Phys. Rev. Lett.}\ }\textbf {\bibinfo {volume} {110}},\ \bibinfo
  {pages} {086401} (\bibinfo {year} {2013})}\BibitemShut {NoStop}%
\end{thebibliography}%
\end{document}